\documentclass[prb,nofootinbib,twocolumn,superscriptaddress]{revtex4} 
\usepackage{graphicx}% Include figure files
\usepackage{dcolumn}% Align table columns on decimal point
\usepackage{bm}% bold math 
\usepackage{threeparttable}
\usepackage{times}
\usepackage{mathptmx}
\usepackage{lscape}
\usepackage{natbib}
\usepackage{amsmath}
\usepackage{amssymb}
\usepackage{braket}
\usepackage{comment}
\usepackage{color}

\def\degree{\kern-.2em\r{}\kern-.3em}

\begin{document}

%\preprint{APS/123-QED} 

\title{  Tropical Diagram for Linear-Nonlinear Boundary in Canonical Ensemble  }
   
\author{Subaru Sugie}
\affiliation{
Department of Materials Science and Engineering,  Kyoto University, Sakyo, Kyoto 606-8501, Japan\\
}%

\author{Koretaka Yuge}
\affiliation{
Department of Materials Science and Engineering,  Kyoto University, Sakyo, Kyoto 606-8501, Japan\\
}%

\begin{abstract}
{ 
%%\begin{itemize}
%%\item 
For classical discrete systems under constant composition, we re-examine how linear-nonlinear boundary in canonical ensemble, connecting a set of potential energy surface and that of microscopic configuration in thermodynamic equilibrium, is characterized by underlying lattice, from tropical geometry. We here show that by applying suitable tropical limit and multiple coordinate transform to time evolution of discrete dynamical system, reflecting geometric aspect of the nonlinearity, we successfully construct \textit{tropical diagram} caputuring the universal character of linear/nonlinear region on configuration space for $f=2$ structural degree of freedoms (SDFs). The diagram indicates that the  boundary near disordered state is mainly dominated by constraints to individual SDF, while \textit{quasi} linear-nonlinear boundary apart from disordered state is dominated by local non-separability in SDFs. 
  }
\end{abstract}

%\pacs{81.90.+c \sep 61.05.-a \sep 05.20.Gg \sep 05.10.-a \sep 02.30.Zz }

\maketitle

\section{Introduction}
%\begin{itemize}
%\item 
When we consider microscopic configuration (with prepared coordination $\left\{ q_{1},\cdots, q_{f} \right\}$) in thermodynamic equilibrium for classical discrete systems under constant composition, it can be typically obtained through canonical average,
\begin{eqnarray}
\Braket{q_{k}}_{z} = Z^{-1} \sum_{Q} E\left( Q \right)\exp\left( -\beta U\left( Q \right)\right),
\end{eqnarray}
where $Z$ denotes partition function, summation is taken over possible configuration $Q$, and potential energy is exactly given under basis 
$\left\{ q_{i} \right\}$ of generalized Ising model (GIM)\cite{gim} by
\begin{eqnarray}
U\left( Q \right) = \sum_{i=1}^{f} \Braket{U|q_{i}}q_{i}\left( Q \right),
\end{eqnarray}
where $f$ denotes number of SDFs in the system, and $\Braket{\quad|\quad}$ represent inner product, i.e., trace over configurations. 
From above equations, canonical average can be interpreted as a map $\phi$, providing
\begin{eqnarray}
\phi:\quad \left\{ \Braket{U|q_{i}} | i=1,\cdots, f\right\} \mapsto \left\{ \Braket{q_{i}}_{Z} | i=1,\cdots, f \right\},
\end{eqnarray}
which typically exhibits complicated nonlinearity. 
%Since possible configurations astronomically increases with increase of the system size, 
Although various theoretical techniques have been developed to predict equilibrium properties, including Metropolis algorism, entropic sampling and Wang-Landau sampling,\cite{mc1,mc2,mc3,wl} origin of the nonlinearity for $\phi$, in terms especially of lattice geometry, has not been sufficiently understood so far. 

Our recent study found that the nonlinearity can be characterized by newly-introduced vector field $A$ on configuration space, ``anharmonicity in structural degree of freedom (ASDF)'',\cite{asdf} independent of temperature and interactions, which depends only on configurational geometry. To further clarify the complicated behavior of $A$, we have applied tropical geometry to its time evolution for $f=1$ SDF system, indicating that magnitude of spatial constraint to the SDF dominate linear-nonlinear boundary.\cite{tropy} 
Since the study was restricted to $f=1$ SDF, it has been still unclear whether or not constraints specific to multiple SDFs, e.g., non-separable character in SDFs, affects the nonlinearity. 
We here tuckle this problem by constructing tropical diagram characterizing linear-nonlinear boundary for a minimal $f=2$ SDF system, based on tropical geometry with specific, multiple coordinate transform. The result certainly exhibit that there exists two dominant contribution to the nonlinearity, from constraints to individual SDF and specific to multiple SDFs, which depends on magnitude of ordering.  The details are shown below.

\section{Derivation and Applications}
\subsection{Description of Nonlinearity in $\phi$}
Let us first briefly introduce basis functions descriping microscopic configuration and potential energy surface, and vector field of ASDF that will be applied to tropical limit disccussed later. 
We here employ GIM on discrete binary X-Y system, where occupation of given lattice site $i$ is specified by spin variable $\sigma_{i}=1$ for X and $\sigma_{i}=-1$ for Y. Then, microscopic configuration $Q=\left( q_{1}, \cdots, q_{f} \right)$ along chosen $k$-th basis function is given by
\begin{eqnarray}
q_{k}\left( Q \right) = \Braket{ \prod_{i\in S_{k}} \sigma_{i} }_{Q},
\end{eqnarray}
where $\Braket{\quad }_{Q}$ denotes taking linear average for configuration $Q$, and the product is taken over a set of symmetry-equivalent figure $S_{k}$ of $k$. Hereinafter, withoug lack of generality, we describe configuration $Q$ measured from center of gravity for configurational density of states (CDOS), which is the density of states for possible configurations before applying many-body interaction to the system.
Under these preparations, ASDF can be defined as vector field, namely
\begin{eqnarray}
A\left( Q \right) = \left\{ \phi\left( \beta \right) \circ \left( -\beta\cdot \Lambda \right)^{-1} \right\} \cdot Q - Q,
\end{eqnarray}
where $\Lambda$ is an $f\times f$ real symmetric matrix of 
\begin{eqnarray}
\Lambda_{ik} = \sqrt{\frac{\pi}{2}}\Braket{q_{k}}_{2} \Braket{q_{i}}_{k}^{\left( + \right)},
\end{eqnarray}
where $\Braket{\quad}_{2}$ denotes taking standard deviation for CDOS, and 
$\Braket{q}_{k}^{\left( + \right)}$ denotes taking linear average over configuration satisfying $q_{k}\ge 0$. 
When $\phi$ is a linear map, $A$ takes zero vector, and $A$ is independent of temperature and interaction, depending only on landscape of CDOS. Therefore, $A$ is a natural measure of nonlinearity in $\phi$, whose successive evolution can be interpreted as the following discrete dynamical system:
\begin{eqnarray}
Q_{a+1} = Q_{a} + A\left( Q_{a} \right).
\end{eqnarray}

In a similar fashion to $f=1$ SDF system, we woule like to start seeing how information about nonlinearity in $\phi$ around origin of CDOS (i.e., perfectly disordered state) evolves when configuration $Q$ aparts from the origin. Here we focus on equiatomic binary system with $r$-th and $s$-th pair coordination of $Q_{r}$ and $Q_{s}$.  Then we can perform series expansion of the vector field $A=\left( A_{r}, A_{s} \right)$ with moments, namely
\begin{eqnarray}
  A_r\left( Q_{r}, Q_{s} \right) \simeq \sum_{k=2}^{\infty} \bigg[\sum_{l=0}^{k} \dfrac{l+1}{D_r} J(k+1,l+1) (Q_{r})^l(Q_{s})^{k-l} \bigg],
\end{eqnarray}
where $D_{r}$ denotes number of $r$-th pair per site, and $J\left( k,l \right)$ denotes number of simple cycles per site consisting of $l$ $r$-th pair and $\left( k-l \right)$ $s$-th pair. %Figure~\ref{fig:cycle} illustrates simulated value of $l\cdot J\left( k,l \right)/D_{r}$ for fcc equiatomic binary system, which shows that this can be approximated by 
Figure for simple cycle illustrates simulated value of $l\cdot J\left( k,l \right)/D_{r}$ for fcc equiatomic binary system, which shows that this can be approximated by 
\begin{eqnarray}
\label{eq:plane}
\frac{\left( l+1 \right)}{D_{r} } J\left( k+1,l+1 \right)\simeq \exp\left\{ M_{1}\left( l+1 \right)+ M_{2}\left( k-l \right) \right\},
\end{eqnarray}
%where $M_{1}$ and $M_{2}$ are the constant depending only on pair type (in Fig.~\ref{fig:cycle}, 
where $M_{1}$ and $M_{2}$ are the constant depending only on pair type,  
$M_{1}\simeq 2.0$ and $M_{2}\simeq 1.5$). 
We confirm that Eq.~\eqref{eq:plane} holds for other representative combination of pairs on fcc lattice, and hereinafter we employ expression of Eq.~\eqref{eq:plane} for number of simple cycles. 
The approximation in Eq.~\eqref{eq:plane} for number of simple cycles in exponential form can be naturally adopted in terms of upper bound for $\log J$, simply given by $\log J\left( k,l \right) \le k \log \left( 4 D_{r} \right) + l \log \left( 4D_{l} \right)$ (see Appendix).

In order to apply tropical limit\cite{trop} to $A_{r}$, i.e., $\lim_{t\to\infty}A_{r}$, we should first employ appropriate coordinate transform because lower and upper bound of GIM basis function for binary system respectively takes -1 and 1. In a similar fashion to $f=1$ SDF system,\cite{tropy} we here employ  
coordinate transform of $Q'_{r}=Q_{r}+1$ and $Q'_{s}=Q_{s}+1$, leading to
\begin{widetext}
\begin{eqnarray}
  A_r\left( Q'_{r},Q'_{s} \right) &=& \sum_{k=2}^{\infty} \bigg[ e^{M_1+M_2k} \sum_{l=0}^{k} e^{\Delta M l} \bigg(\sum_{n=0}^{l} {}_{l} C_n (Q'_{r})^{n}(-1)^{l-n}\bigg) \bigg(\sum_{m=0}^{k-l} {}_{k-l} C_m (Q'_{s})^m (-1)^{k-l-m} \bigg)  \bigg],
\end{eqnarray}
\end{widetext}
where $\Delta M \equiv M_{1} - M_{2}$. In the above equation, coefficient $C_{u,v}$ for $\left( Q'_{r} \right)^{u}\left( Q'_{s} \right)^{v}$ can be further simplified to (see Appendix)
\begin{widetext}
\begin{eqnarray}
\label{eq:coeff}
C_{u,v} = e^{M_1}\cdot (-1)^{u+v} \lim_{n \rightarrow \infty} \bigg[\sum_{k=u}^{n} {}_k C_v (-e^{M_1})^k\bigg] \lim_{m \rightarrow \infty} \bigg[\sum_{k=v}^{m} {}_k C_u (-e^{M_2})^k\bigg].
\end{eqnarray}
\end{widetext}
For large $k$, terms in bracket of each limit in \eqref{eq:coeff} can be approximated by geometric sequence in analogy to 1-SDF system,\cite{tropy} which leads to 
\begin{widetext}
\begin{eqnarray}
\label{eq:afin}
A_{r}\left( Q'_{r},Q'_{s} \right) &=&  \dfrac{e^{M_1}}{(1+e^{M_1})(1+e^{M_2})} \cdot \lim_{n \rightarrow \infty} \bigg[e^{(n+1)M_1}(-1)^n \sum_{v=0}^{\alpha} {}_n C_v (Q_r')^{v}\bigg] \cdot \lim_{m \rightarrow \infty} \bigg[e^{(m+1)M_2} (-1)^m \sum_{u=0}^{\beta}{}_m C_u  (Q_s')^{u} \bigg] \nonumber \\
&=&\dfrac{e^{2M_1+M_2}}{(1+e^{M_1})(1+e^{M_2})}\lim_{\substack{n\rightarrow\infty \\ m\rightarrow\infty}}\bigg[(-1)^{n+m} \dfrac{[1-(-q_r')^{\alpha+1}][1-(-q_s')^{\beta+1}]}{(1+q_r')(1+q_s')}\bigg]
\end{eqnarray}
\end{widetext}
under the conditions for $\alpha$ and $\beta$ of  
\begin{eqnarray}
\label{eq:cond}
\lim_{n\rightarrow \infty} \dfrac{{}_n C_{\alpha}}{e^{nM_1}} &=& 0  \nonumber \\
\lim_{m\rightarrow \infty} \dfrac{{}_m C_{\beta}}{e^{mM_2}} &=& 0,
\end{eqnarray}
which means taking higher maximum degree of simple cycles than maximum powers of $Q'_{r}$ and $Q'_{s}$. 
We note here that by simply applying tropical limit to Eq.\eqref{eq:afin}, geometric information of lattice is almost disappeared. Since we take $n\to\infty$ and $m\to\infty$, we can avoid this problem by explicitly consider $e^{n}$ and $e^{m}$ as additional variable, where
\begin{eqnarray}
\log_{t}e^{n} &=& \mu \nonumber \\
\log_{t}e^{m}&=& \lambda.
\end{eqnarray}
Neccessary conditions for $\mu$ and $\lambda$ would be naturally determined so that at origin, tropical limit of $A$ takes zero (i.e., $\phi$ provides locally linear map at perfectly disordered state, which should hold for any systems). 

Under these preparations, we are now ready to taking tropical limit for Eq.~\eqref{eq:afin}. 
For convenience, we first define the followings:
\begin{widetext}
\begin{eqnarray}
\omega_{r,a} &=& \lim_{t\to\infty} Q'_{r,a} , \quad \omega_{s,a} = \lim_{t\to\infty} Q'_{s,a}, \quad M_r = M_1 \mu + M_2 \lambda,\quad M_s = M_1 \lambda + M_2 \mu \nonumber \\
S_{r,a+1} &=& Q_{r,a+1}'(1+Q_{r,a}')(1+Q_{s,a}'), \quad S_{r,a} = Q_{r,a}'(1+Q_{r,a}')(1+Q_{s,a}') \nonumber \\
%T_{r,a+1} &=& \max{\{\omega_{r,a+1},\omega_{r,a+1}+\omega_{r,a},\omega_{r,a+1}+\omega_{s,a},\omega_{r,a+1}+\omega_{r,a}+\omega_{s,a}\}} \nonumber \\
T_{r,a} &=& \max{\{\omega_{r,a},2\omega_{r,a},\omega_{r,a}+\omega_{s,a},2\omega_{r,a}+\omega_{s,a}\}}.
\end{eqnarray}
\end{widetext}
Then, we should individually consider tropical limit of Eq.~\eqref{eq:afin} for (i) parity of $m$ and $n$ and (ii) signs of $\omega_{r,a}$ and $\omega_{s,a}$, showing in the followings.

\subsection{Tropical Limit of Nonlinearity}
\subsubsection{$m=$even, $n=$even}
In this case, Eq.~\eqref{eq:afin} can be rewritten as 
\begin{eqnarray}
&&S_{r,a+1} + H[q_{s,a}'^{\beta+1} + q_{r,a}'^{\alpha+1} q_{s,a}'^{\beta+1}]e^{nM_1 + (m+1)M_2} \notag \\
&=& S_{r,a} + H [1 + q_{r,a}'^{\alpha+1}}]e^{nM_1 + (m+1)M_2.
\end{eqnarray}
The corresponding toropical limit takes
\begin{widetext}
\begin{eqnarray}
\label{eq:tr1}
&&\mathrm{\left( i \right)}\omega_{r,a} > 0,  \omega_{s,a} > 0 \nonumber \\
  &&\quad\quad\max\{\omega_{r,a+1} + \omega_{r,a} + \omega_{s,a} ,M_r + (\alpha+1) \omega_{r,a} + (\beta+1) \omega_{s,a}\} = \max\{2\omega_{r,a} + \omega_{s,a},M_r+(\alpha+1)\omega_{r,a}\} \nonumber \\
&&\mathrm{\left( ii \right)}\omega_{r,a} > 0,  \omega_{s,a} \le 0 \nonumber \\
  &&\quad\quad\max \{\omega_{r,a+1} + \omega_{r,a},M_r + (\alpha+1)\omega_{r,a} + (\beta+1)\omega_{s,a}\} = \max \{2\omega_{r,a},M_r + (\alpha+1) \omega_{r,a}\} \nonumber \\
&&\mathrm{\left( iii \right)}\omega_{r,a} \le 0,  \omega_{s,a} > 0 \nonumber \\
  &&\quad\quad\max \{\omega_{r,a+1} + \omega_{s,a} ,M_r + (\beta+1)\omega_{s,a}\} = \max \{\omega_{r,a} + \omega_{s,a}, M_r \} \nonumber \\
&&\mathrm{\left( iv \right)}\omega_{r,a} \le 0,  \omega_{s,a} le 0 \nonumber \\
&&\quad\quad\max \{\omega_{r,a+1},M_r+(\beta+1)\omega_{s,a}\} = \max \{\omega_{r,a},M_r\}
\end{eqnarray}
\end{widetext}

\subsubsection{$m=$even, $n=$odd}
Eq.~\eqref{eq:afin} can be rewritten as
\begin{eqnarray}
 &&S_{r,a+1} + H[1 +  q_{r,a}'^{\alpha+1} q_{s,a}'^{\beta+1}] e^{nM_1 + (m+1)M_2} \notag \\
  &=& S_{r,a} + H[q_{r,a}'^{\alpha+1} + q_{s,a}'^{\beta+1}] e^{nM_1 + (m+1)M_2},
\end{eqnarray}
leading to the tropical limit of
\begin{widetext}
\begin{eqnarray}
\label{eq:tr2}
&&\mathrm{\left( i \right)}\omega_{r,a} > 0,  \omega_{s,a} > 0 \nonumber \\
  &&\quad\quad \max\{\omega_{r,a+1}+\omega_{r,a} + \omega_{s,a} ,M_r + (\alpha+1) \omega_{r,a} + (\beta+1) \omega_{s,a}\} = \max\{2\omega_{r,a} + \omega_{s,a},M_r+(\alpha+1)\omega_{r,a},M_r + (\beta+1)\omega_{s,a}\} \nonumber \\
&&\mathrm{\left( ii \right)}\omega_{r,a} > 0,  \omega_{s,a} \le 0 \nonumber \\
  &&\quad\quad \max \{\omega_{r,a+1}+\omega_{r,a},M_r,M_r + (\alpha+1)\omega_{r,a} + (\beta+1)\omega_{s,a}\} = \max \{2\omega_{r,a},M_r + (\alpha+1) \omega_{r,a}\} \nonumber \\
&&\mathrm{\left( iii \right)}\omega_{r,a} \le 0,  \omega_{s,a} > 0 \nonumber \\
  &&\quad\quad \max \{\omega_{r,a+1} + \omega_{s,a} ,M_r ,M_r + (\alpha+1)\omega_{r,a}+ (\beta+1)\omega_{s,a}\} = \max \{\omega_{r,a} + \omega_{s,a}, M_r + (\beta+1) \omega_{s,a} \} \nonumber \\
&&\mathrm{\left( iv \right)}\omega_{r,a} \le 0,  \omega_{s,a} le 0 \nonumber \\
  &&\quad\quad \max \{\omega_{r,a+1},M_r\} = \max \{\omega_{r,a},M_r+(\alpha+1)\omega_{r,a},M_r + (\beta+1) \omega_{s,a}\}.
\end{eqnarray}
\end{widetext}

\subsubsection{$m=$odd, $n=$even}
This condition rewrites Eq.~\eqref{eq:afin} as
\begin{eqnarray}
  &&S_{r,a+1} + H[1 + q_{r,a}'^{\alpha+1} + q_{s,a}'^{\beta+1}+q_{r,a}'^{\alpha+1} q_{s,a}'^{\beta+1}]e^{nM_1 + (m+1)M_2} \notag \\
  &=&S_{r,a}, 
\end{eqnarray}
leading to the tropical limit of
\begin{widetext}
\begin{eqnarray}
\label{eq:tr3}
&&\mathrm{\left( i \right)}\omega_{r,a} > 0,  \omega_{s,a} > 0 \nonumber \\
  &&\quad\quad \max\{\omega_{r,a+1}+\omega_{r,a} + \omega_{s,a} ,M_r + (\alpha+1)\omega_{r,a} + (\beta+1)\omega_{s,a}\} = 2\omega_{r,a} + \omega_{s,a} \nonumber \\
&&\mathrm{\left( ii \right)}\omega_{r,a} > 0,  \omega_{s,a} \le 0 \nonumber \\
  &&\quad\quad \max \{\omega_{r,a+1} + \omega_{r,a},M_r + (\alpha+1)\omega_{r,a}\} = 2\omega_{r,a}  \nonumber \\
&&\mathrm{\left( iii \right)}\omega_{r,a} \le 0,  \omega_{s,a} > 0 \nonumber \\
  &&\quad\quad \max \{\omega_{r,a+1} + \omega_{s,a} ,M_r  + (\beta+1)\omega_{s,a}\} = \omega_{r,a} + \omega_{s,a}  \nonumber \\
&&\mathrm{\left( iv \right)}\omega_{r,a} \le 0,  \omega_{s,a} le 0 \nonumber \\
  &&\quad\quad \max \{\omega_{r,a+1},M_r\} = \omega_{r,a}
\end{eqnarray}
\end{widetext}

\subsubsection{$m=$odd, $n=$odd}
We finally consider the rest condition for the parity of $m$ and $n$, providing Eq~\eqref{eq:afin} as
\begin{eqnarray}
  && S_{r,a+1} + H[q_{r,a}'^{\alpha+1} + q_{r,a}'^{\alpha+1} q_{s,a}'^{\beta+1}]e^{nM_1 + (m+1)M_2} \notag \\
  &=& S_{r,a} + H[1 + q_{s,a}'^{\beta+1}]e^{nM_1 + (m+1)M_2}.
\end{eqnarray}
This leads to the tropical limit of
\begin{widetext}
\begin{eqnarray}
\label{eq:tr4}
&&\mathrm{\left( i \right)}\omega_{r,a} > 0,  \omega_{s,a} > 0 \nonumber \\
  &&\quad\quad \max\{\omega_{r,a+1}+\omega_{r,a} + \omega_{s,a} ,M_r + (\alpha+1) \omega_{r,a} + (\beta+1) \omega_{s,a}\} = \max\{2\omega_{r,a} + \omega_{s,a},M_r+(\beta+1)\omega_{s,a}\} \nonumber \\
&&\mathrm{\left( ii \right)}\omega_{r,a} > 0,  \omega_{s,a} \le 0 \nonumber \\
  &&\quad\quad  \max \{\omega_{r,a+1} + \omega_{r,a},M_r + (\alpha+1)\omega_{r,a}\} = \max \{2\omega_{r,a},M_r \} \nonumber \\
&&\mathrm{\left( iii \right)}\omega_{r,a} \le 0,  \omega_{s,a} > 0 \nonumber \\
  &&\quad\quad  \max \{\omega_{r,a+1} + \omega_{s,a} ,M_r + (\alpha+1)\omega_{r,a} + (\beta+1)\omega_{s,a}\} = \max \{\omega_{r,a} + \omega_{s,a}, M_r + (\beta+1)\omega_{s,a}\} \nonumber \\
&&\mathrm{\left( iv \right)}\omega_{r,a} \le 0,  \omega_{s,a} le 0 \nonumber \\
  &&\quad\quad \max \{\omega_{r,a+1},M_r+(\alpha+1)\omega_{r,a}\} = \max \{\omega_{r,a},M_r\}
\end{eqnarray}
\end{widetext}

\subsection{Overall Behavior for Tropical Limit of Nonlinearity}
\begin{figure}[h]
\begin{center}
\includegraphics[width=1.03\linewidth]{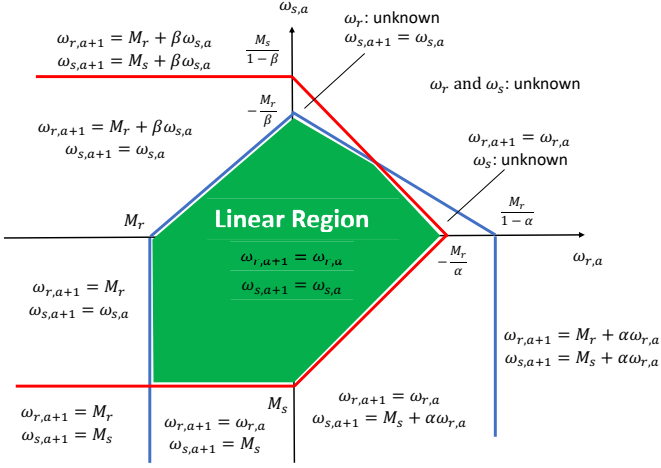}
\caption{Nonlinearity in $\phi$  in terms of tropical limit of ASDF.  } 
\label{fig:diag}
\end{center}
\end{figure}

By combining the results of Eqs.~\eqref{eq:tr1}-\eqref{eq:tr4}, we can illustrate diagram for tropical limit of nonlinearity in $\phi$, as shown in 
Fig.~\ref{fig:diag}. 
Here, we impose condition of $M_{r}<0$ and $M_{s}<0$ to retain linear character at the origin, which always holds for any discrete systems in terms of the original dynamical system.
Note that in Eqs.~\eqref{eq:tr1}-\eqref{eq:tr4}, all combination of parity of $m$ and $n$ results in the same tropical limit of ASDF, except for the difference in quadrant where information of ASDF loses. From Fig.~\ref{fig:diag}, several important features can be clearly seen: (i) At the third quadrant (i.e., ordering tendency), linear region is independently bounded by individual SDF of $M_{r}$ and $M_{s}$, while at the first quadrand (i.e., clustering tendency), it is bounded through the \textit{interaction} between SDFs since gradient of the boundary is affected by both SDF-originated information of $M_{r}$, $M_{s}$, $\alpha$ and $\beta$. (ii) Partial information about ASDF loses outside of the linear region at the first quadrant (described by ``unknown'' in Fig.~\ref{fig:diag}). For these two features, we emphasize here that the first quadrant of Fig.~\ref{fig:diag} corresponds to tropical limit for $\left\{ 1<Q'_{r}, 1<Q'_{s} \right\}$ while the third quadrant for $\left\{ 0<Q'_{r}<1, 0<Q'_{s}<1 \right\}$, i.e., the former is semi-infinite area and the latter is finite area: Therefore, we should examine whether or not the difference in boundary for linear region at the first and the third quadrant comes from finite/infinite area difference.
In order to confirm this, we first apply the following coordinate transform of 180 degree rotation, namely
\begin{eqnarray}
P'_{r} = -Q_{r} +1,\quad P'_{s} = -Q_{s} + 1.
\end{eqnarray}

\begin{figure}[h]
\begin{center}
\includegraphics[width=1.03\linewidth]{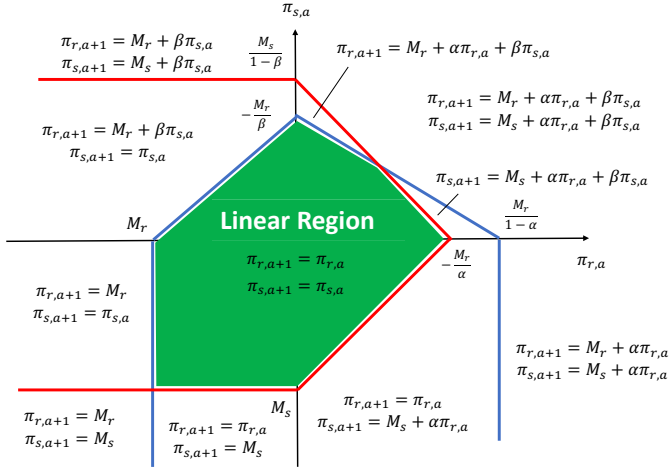}
\caption{ Nonlinearity in $\phi$  in terms of tropical limit of ASDF after coordinate transform of 180 degree roration.}
\label{fig:diag2}
\end{center}
\end{figure}
In a similar fashion for the case of original coordinate, the corresponding series expantion leads to 
\begin{widetext}
\begin{eqnarray}
  A_r(P_r',P_s') &=&\sum_{k=2}^{\infty} \bigg\{\sum_{l=0}^k \bigg[e^{M_1(k-l)+M_2(l+1)} \bigg( \sum_{n=0}^l  {}_n C_u (-P_r')^n \bigg) \bigg( \sum_{m=0}^{k-l} {}_{k-l} C_m (-P_s')^m \bigg) \bigg] \bigg\} \nonumber \\
 &\simeq& \dfrac{e^{M_1+2M_2}}{(1+e^{M_1})(1+e^{M_2})} \cdot e^{nM_1+mM_2} \cdot(-1)^{n+m} \cdot \dfrac{[1-(-P_r')^{\alpha+1}][1-(-P_s')^{\beta+1}]}{(1+P_r')(1+P_s')},
\end{eqnarray}
\end{widetext}
which is now ready for applying the tropical limit. 
We also define the followings:
\begin{eqnarray}
\label{eq:trans}
\pi_{r,a} &=& \lim_{t\to\infty} P'_{r,a} , \quad \pi_{s,a} = \lim_{t\to\infty} P'_{s,a}.
\end{eqnarray}
The resultant diagram for tropical limit of nonliearity is summarized in Fig.~\ref{fig:diag2}.
Figure~\ref{fig:diag2} shows important features of (i) information about nonliear region in the third quadrant (corresponding to the first quadrant for the original coordinate) appears, and (ii) the linear region is independent of coordinate transform. 
These facts strongly indicate that for original coordinate, the difference in boundary for linear region at the first and the third quadrant mainly comes from difference in tropical limit for finite and infinite area: Near the origin (i.e., perfectly disordered state), linear character of $\phi$ is mainly bounded by constraints to individual SDF both for ordering (at the third quadrant of Fig.~\ref{fig:diag}) and clustering (at the third quadrant of Fig.~\ref{fig:diag2}) configuration, and the ``interaction'' between SDFs becomes dominant role for bounding linear region when given ordering or clustering configuration is far from the origin (i.e., the first quadrants of Figs.~\ref{fig:diag} and~\ref{fig:diag2}).
This indicates that by applying simple coordinate transform, we can selectively extract dominant contribution to linear/nonlinear boundary at near or far from the disordered state, respectively appears at the third and the first quadrant both before and after applying coordinate transform. The difference in linear region boundary can be naturally accepted from the fact that for configurations far from the origin, configurational polyhedra (its vertex corresponds to farthest from origin) strongly depends on lattice and combination of SDFs, which reflects explicit constraints to SDF by underlying lattice: The interactions in SDFs can be related to non-separable character in CDOS.
Very recently, we have successfully introduce another measure for the local nonlinearity in terms of the ASDF,\cite{ig} which is based on the Kullback-Leibler divergence on statistical manifold: The measure can explicitly decompose the non-separable contribution in SDFs from local nonlinearity,  which will enable us to further quantify the difference in the boundary for linear region, addressed in our future study.

\section{Conclusions}
For classical discrete systems under constant composition, we theoretically address how linearity around perfectly disordered state is bounded in terms of structural degree of freedoms, based on tropical geometry. We show that by applying suitable tropical limit and multiple coordinate transform for evolution of the previously-introduced special vector field reflecting local nonlinearity, we can selectively see linear/nonlinear boundary near or far from the disordered state, respectively dominated by constraints to individual SDF and by interaction between SDFs, which can be the main role for lattice as constraints to SDFs.

\section{Acknowledgement}
This work was supported by Grant-in-Aids for Scientific Research on Innovative Areas on High Entropy Alloys through the grant number JP18H05453 and a Grant-in-Aid for Scientific Research (16K06704) from the MEXT of Japan, Research Grant from Hitachi Metals$\cdot$Materials Science Foundation.

\appendix
\section*{Appendix}
\subsection{Upper bound for number of simple cycles}
Let us consider pair figure $r$ and $s$ where their coordination number is given by $D_{r}$ and $D_{s}$. Then it is clear that number of simple cycles $J\left( n_{r}, n_{s} \right)$ consisting of $n_{r}$ $r$-pairs and $n_{s}$ $s$-pairs satisfies the following inequality:
\begin{eqnarray}
\label{eq:sap}
  J_{\rm fcc}(n_r,n_s) < {}_{n_r+n_s}C_{n_r}(D_r)^{n_r}(D_s)^{n_s}.
\end{eqnarray}
When we apply Stirling's formula with $n_{r} + n_{s} = 2N$, we obtain
\begin{eqnarray}
  {}_{n_r + n_s} C_{n_r} &=& {}_{2N} C_{n_r} \leq {}_{2N} C_N = \dfrac{(2N)!}{N!N!} \nonumber \\
  \log{({}_{2N} C_N)} &\simeq& (2N\log{2N} - 2N) - 2(N\log{N} - N) \nonumber \\
  &=& 2N \log{2N} - 2N \log{N} \nonumber \\
  &=& 2N \log{2}.
\end{eqnarray}
Therefore, we can rewrite Eq.~\eqref{eq:sap} as 
\begin{eqnarray}
  \log{J_{\rm fcc}(n_r,n_s)} &<& \log{({}_{n_r+n_s} C_{n_r}(D_r)^{n_r}(D_s)^{n_s})} \nonumber \\
  &\leq& 2(n_r+n_s)\log{2} + n_r \log{D_r} + n_s \log{D_s} \nonumber \\ 
  &=& n_r\log{(4D_r)} + n_s\log{(4D_s)}.
\end{eqnarray}

\subsection{Derivation of Eq.~\eqref{eq:coeff}}
We here consider the condition of $k=u+v+i(i\geq 0)$, which can rewrite coefficient of 
\begin{widetext}
\begin{eqnarray}
  C_{u,v} = \sum_{i=0}^{\infty} e^{M_2(u+v+i)+M_1} \bigg[\sum_{t=0}^{i} e^{\Delta M(u+t)} {}_{u+t} C_u (-1)^t \cdot {}_{v+(i-t)} C_v (-1)^{i-t} \bigg].
\end{eqnarray}
\end{widetext}
This can be further transformed into
\begin{widetext}
\begin{eqnarray}
  &&\dfrac{C_{u,v}}{e^{M_2(u+v)+M_1}} =  e^{M_1 \cdot 0} \bigg[e^{\Delta M u} {}_u C_u (-1)^0 {}_v C_v (-1)^0 \bigg] \notag \\
  && +e^{M_1 \cdot 1} \bigg[e^{\Delta M u} {}_u C_u (-1)^0 {}_{v+1} C_v (-1)^1 + e^{\Delta M(u+1)} {}_{u+1} C_u(-1)^1 {}_v C_v (-1)^0 \bigg] \\
  && +e^{M_1 \cdot 2} \bigg[e^{\Delta M u} {}_u C_u (-1)^0 {}_{v+2} C_v (-1)^2 + e^{\Delta M(u+1)} {}_{u+1} C_u (-1)^1 {}_{v+1} C_v (-1)^1 \notag\\
  &&+e^{\Delta M(u+2)} {}_{u+2} C_u (-1)^2 {}_v C_v (-1)^0 \bigg] \notag \\
  && +\cdots \\
  &&= e^{\Delta M u} {}_u C_u (-1)^0 \bigg[{}_v C_v (-e^{M_1})^0 + {}_{v+1} C_v (-e^{M_1})^1 + {}_{v+2} C_v (-e^{M_1})^2 + \cdots \bigg] \notag \\
  &&+ e^{\Delta M(u+1)} {}_{u+1} C_u (-e^{M_1}) \bigg[{}_v C_v (-e^{M_1})^0 + {}_{v+1} C_v (-e^{M_1})^1 + {}_{v+2} C_v (-e^{M_1})^2 + \cdots \bigg]  \notag \\
  &&+ e^{\Delta M(u+2)} {}_{u+2} C_u (-e^{M_1})^2 \bigg[{}_v C_v (-e^{M_1})^0 + {}_{v+1} C_v (-e^{M_1})^1 + {}_{v+2} C_v (-e^{M_1})^2 + \cdots \bigg] \notag \\
  &&+ \cdots.
\end{eqnarray}
\end{widetext}
When we introduce 
\begin{eqnarray}
  K &\equiv& \sum_{n=v}^{\infty} {}_n C_v (-e^{M_1})^{n-v}, \nonumber \\
  L &\equiv& \sum_{m=u}^{\infty} {}_m C_u (-e^{M_2})^{m-u}, 
\end{eqnarray}
we can obtain the following since $e^{\Delta M u}\cdot e^{M_2(u+v)} = e^{M_1u + M_2v}$, namely
\begin{widetext}
\begin{eqnarray}
  C_{u,v} &=&K \cdot e^{M_1(u+1)+M_2v} \cdot \bigg[{}_u C_u (-e^{M_2})^0 + {}_{u+1} C_u (-e^{M_2})^1 + {}_{u+2} C_u (-e^{M_2})^2 + \cdots \bigg] \nonumber \\
  &=& e^{M_1(u+1)+M_2v} K L . \nonumber \\
  &=& e^{M_1} \cdot  (-1)^{u+v} \bigg[ \sum_{n=v}^{\infty} {}_n C_v (-e^{M_1})^n\bigg] \bigg[\sum_{m=u}^{\infty}{}_mC_u (-e^{M_2})^m\bigg] \nonumber \\
  &=& e^{M_1}\cdot (-1)^{u+v} \lim_{n \rightarrow \infty} \bigg[\sum_{k=u}^{n} {}_k C_v (-e^{M_1})^k\bigg] \lim_{m \rightarrow \infty} \bigg[\sum_{k=v}^{m} {}_k C_u (-e^{M_2})^k\bigg].
\end{eqnarray}
\end{widetext}

\end{document}